\documentclass[preprint]{elsarticle}
\pdfoutput=1 % if your are submitting a pdflatex (i.e. if you have
\usepackage{amsmath}
\usepackage{caption}
\usepackage{graphicx,xcolor}      % alternative graphics specifications
\usepackage{longtable, gensymb}     % helps with long table options
\usepackage{url,physics}           % for on-line citations
\usepackage{bm}            % special 'bold-math' package
\usepackage[english]{babel}
\usepackage{ulem}

\newcommand*\bob{\color{black}}

\begin{document}
\title{Antimatter as Macroscopic Dark Matter}

\author[1]{Jagjit Singh Sidhu\corref{cor1}}
\ead{jxs1325@case.edu}
\author[2]{Robert J. Scherrer}
\ead{robert.scherrer@vanderbilt.edu}
\author[1]{Glenn Starkman}
\ead{glenn.starkman@case.edu}
\cortext[cor1]{Corresponding author}

\address[1]{Physics Department/CERCA/ISO Case Western Reserve University  \\
	Cleveland, Ohio 44106-7079, USA}
\address[2]{Department of Physics $\&$ Astronomy, Vanderbilt University,
Nashville, TN 37235}
\iffalse
\author[a]{Jagjit Singh Sidhu}
\author[b]{Robert Scherrer}
\author[a]{Glenn Starkman}

\affiliation[a]{Physics Department/CERCA/ISO Case Western Reserve University  \\
	Cleveland, Ohio 44106-7079, USA}
\affiliation[b]{Department of Physics $\&$ Astronomy, Vanderbilt University,
Nashville, TN 37235 }
\emailAdd{jxs1325@case.edu}
\emailAdd{robert.scherrer@vanderbilt.edu}
\emailAdd{gds6@case.edu}
\fi

% \date{June 2019}

%\begin{affiliations}
%\item Physics Department/CERCA/ISO Case Western Reserve University Cleveland, Ohio 44106-7079, USA
%\item Department of Physics $\&$ Astronomy, Vanderbilt University,
%Nashville, TN 37235 
%\end{affiliations}

\begin{abstract}
% Dark matter is a vital component of the current best model of our universe $\Lambda$CDM. There are numerous
% leading dark matter candidates, most notably particle
% candidates such as WIMPS and axions, 
% However, a lack of a direct
% detection of these dark matter particle candidates implies
% that composite dark matter objects should receive
% more attention.
Antimatter macroscopic dark matter {\bob (macros)} refers to a generic {\bob class} of 
antimatter dark matter candidates that interact with
ordinary matter primarily through annihilation
with large cross-sections. 
A combination of terrestrial,
astrophysical, and cosmological observations constrain a portion of the anti-macro parameter space. However,
a large region of the parameter space remains
unconstrained, most notably for nuclear-dense
objects. 
%The consequences for a particular
%type of anti-macro objects is discussed in light of
%these constraints.

\end{abstract}

\maketitle

\section{Introduction}

{\bob The evidence for dark matter is overwhelming (see,
e.g., \cite{a} and references therein), but}
the nature of 
dark matter remains one of the great
unsolved mysteries of modern cosmology.
In recent years the authors 
have {\color{black} explored the proposition}
that the dark matter might be macroscopic, in the sense of
having a characteristic mass $M_x$ and cross-sectional area in the gram and cm$^2$ range,
respectively
\cite{b, b1, atmos, q, dbdm, 1908.00557, thermo2, g2}.
The dark matter constituents in this model are called ``macros."
The macro model has two undetermined parameters,
the macro mass $M_x$ and the interaction
cross section $\sigma_x$.
The dominant
interaction is assumed to be
elastic scattering, with
$\sigma_x$
taken
to be the geometric cross-section of the macro.

We begin
by first briefly reviewing the 
existing constraints on macros
derived
in previous works.
For macro
masses $M_x \leq 55\,$g careful examination of specimens of
old mica for tracks made by passing dark matter \cite{b, h,i}
has ruled out such objects as the primary dark-matter
candidate. 
For $M_x \geq 10^{23}\,$g, a variety of microlensing searches have constrained the abundance of
macros \cite{i2,j,k,l,m} from a lack
of magnification of sources
by a passing macro along the line
of sight of the observer. 

A large region of parameter space was constrained by considering thermonuclear runaways triggered by macros incident on white dwarfs 
\cite{thermo1}.
However, it was later shown \cite{thermo2} that
the excluded region of macro parameter space for macros
providing
all of the dark matter was too large, and
more accurate constraints were placed. 
Dark matter-photon elastic interactions 
were used together with the Planck Cosmic
Microwave Background (CMB) data to constrain macros of sufficiently high reduced
cross-section $\sigma_x/M_x$ \cite{Boehm}.
% Prior
% work had already constrained a similar range of parameter space by showing that 
% the consequence of dark matter
% interactions with standard model particles is to dampen
% the primordial matter 
% fluctuations and essentially erase
% all structures below a given scale (see e.g. \cite{Bhm}).
The region of parameter
space where macros would have produced obvious devastating
injuries was also
constrained \cite{dbdm}.

In addition to these constraints, limits from possible future observations have also been proposed. 
Ultra-high-energy cosmic ray detectors that exploit
atmospheric fluorescence could potentially be modified to
probe parts of macro parameter space \cite{atmos}, including
macros of nuclear density. This analysis
has led to constraints being placed
using networks of cameras that were originally built
to study bolides, i.e. extremely bright meteorites
\cite{1908.00557}. Some of us have also suggested
how the approach applied
to mica \cite{h,i} could be applied to a larger, widely available sample of appropriate rock \cite{q}, and used to search for
larger-mass macros. 
In addition to that, we have identified additional
regions of parameter space
constrained by the duration
between back-to-back
superbursts (thermonuclear
runaway on the outer surface of
a neutron star) \cite{thermo2}.

It is unlikely that macro masses beyond $\sim 10^{9}\,$g
could be probed by any purpose-built terrestrial detector assuming even an observation time of a century and
a target area the size of the Earth. Terrestrial probes
(eg. ancient rocks \cite{q,h,i}) could have been continuously exposed for up to $\sim 3 \times 10^9$ years, but we are unlikely
to carefully examine the more than $1\,$km$^2$
that would be
needed to push beyond $M_x \sim 10^9\,$g. It will therefore require innovative thinking about astrophysical probes (eg.
\cite{thermo1}) to probe the
remaining parameter space
at the very highest masses.

%Given the lack of a confirmed direct
%detection for (macroscopic) dark matter,
%it is pertinent to consider the phenomenology
%of an even wider class of such objects than
%previously considered.
In this paper, we consider a related but phenomenologically very different macro model:  the possibility that macros are composed of antimatter. We dub these objects ``anti-macros".  While some macro limits simply carry over to the anti-macro case, we will show that this model yields a rich variety of new phenomena, and we will derive corresponding limits on the anti-macro parameter space.

Anti-macros are most likely
composites of more fundamental anti-particles.
An intriguing possibility is that the anti-macros 
could be made
of Standard Model quarks or baryons bound by Standard Model forces. This suggestion was originally made
by Witten \cite{c}, in the context of a 
first-order QCD phase transition early in the history of the Universe. 
% A
% more realistic version was advanced by Lynn, Nelson and
% Tetradis \cite{d} and Lynn \cite{e}
% in the context of $SU(3)$ chiral perturbation theory.  They
% argued that ``the true bound state
% of nuclei may have two thirds
% of the baryon number consisting
% of strange quarks and that ordinary
% nuclei may only be metastable." 
% Nelson \cite{f} studied the possible
% formation of such ``nuggets of strange baryon matter''
% in an early-universe transition from a kaon-condensate phase of QCD to the ordinary phase.
Others have suggested non-Standard Model versions of such nuclear objects and their formation, for example incorporating the axion  \cite{AQN1,AQN2,AQN3,AQN4,AQN5,AQN6,AQN7,AQN8}. The Axion Quark Nugget (AQN) model
is the most well-studied
model of antimatter macroscopic dark matter in the 
literature. 
These nuggets can be made
of matter as well as antimatter during the QCD phase transition. A 
direct consequence of this feature is that the dark matter and baryon densities 
will automatically assume the same order of magnitude
without any fine tuning.
However, the nuggets in the AQN model
possess a high reflectivity owing to the large potential
of the confining layer of axions \cite{AQN2}. In this manuscript,
we consider a more generic class of anti-macros in which the anti-macro is not bound
by some external layer but by a force
sourced by the anti-macro components themselves.
Hence, we will assume negligible
reflectivity. However, we remain
agnostic about the details of this binding force
and consider only the phenomenology of such objects.
% Within the AQN model,
% the nuggets are expected to possess a distribution of
% masses, with the average baryon number
% of the distribution given by
% \begin{equation}
% \langle B \rangle = \int_{B_{min}}^{B_{max}} B f(B) dB\,,
% \end{equation}
% where $B$ is the baryon number of a quark nugget, 
% $f(B) \propto B^{-\alpha}$ is the expected distribution
% of the nuggets and $\alpha \approx 2 - 2.5\,.$ 

{\bob 
% Once the mass and cross-section
% of the anti-macros are specified, the internal density of an individual macro is completely
% determined by
% the fact that the cross-section is geometric.
Anti-macros corresponding to the models mentioned in the previous paragraph would most likely have densities that are comparable to nuclear density 
(which we take to be $\rho_{nuclear} \approx 3.6 \times 10^{14}\,$g$\,$cm$^{-3}$).
This is much higher than ordinary ``atomic density"
($\rho_{atomic} \approx 1\,$g$\,$cm$^{-3}$), 
but much lower than the density of black holes of masses in the range we consider.
Although anti-macros of approximately nuclear density
are
of particular interest, other densities are not excluded at this point,
so we will consider the full range of possibilities for $M_X$ and $\sigma_X$.}
% Note that 
% anti-macros that form prior to $T \sim 1$ MeV
% are not subject to the bounds on the baryon density
% from Big-Bang nucleosynthesis.}

% {\bob Previous work involving
% anti-macros have generally assumed them
% to be composed of regular anti-baryons 
% (see e.g. \cite{g1} and references therein)
% and to interact primarily through
% elastic scattering with their
% geometric cross-section.
% Additionally, it was shown that
% macros could still
% possess large amounts of electric charge
% and evade detection \cite{g2}.
% Constraints on macro masses and cross sections 
% have always been placed 
% from purely phenomenological
% considerations.}

The sensitivity of a detector to anti-macros
depends on the energy transferred
when the anti-macro transits the detector.
Consider an anti-macro with cross section $\sigma_X$ passing through a detector.
The energy per unit length deposited by a macro 
through annihilation in the detector is 
\begin{equation}\label{dedx}
\frac{dE}{dx} = \kappa \sigma_x \rho c^2,
\end{equation} 
where $\rho$ is the density of the target
and $\kappa$ is introduced, generically,
to account for the fraction of the annihilation energy
that is deposited as heat into the surrounding medium. In
the case of the AQN, 
$\kappa \ll 1$ in most cases of interest, and 
the energy deposition
is highly suppressed.
An order of magnitude
estimate is typically $\kappa \sim 10^{-12}$
(see Appendix C in reference
\cite{AQN7}). For such objects, the energy deposition
is similar to the case of ordinary macros
\cite{b, b1, atmos, q, dbdm, 1908.00557, thermo2, g2}
and the constraints presented in those references apply to both macros and anti-macros.
The energy transfer expression
for ordinary macros is
\begin{equation}\label{macrodedx}
\frac{dE}{dx} = \sigma_x \rho v_x^2\,,
\end{equation} 
where $v_x \sim 250\,$km s$^{-1}$ is the speed of the macro.
Thus, anti-macros with $\kappa \sim 1$ are expected to deposit
$\sim 10^6$ times more energy in a target
than macros of the same cross-section. 
They should, in general, be easier to observe.  The physical reason for this difference is that the anti-macro collisions with ordinary matter convert rest energy into thermal energy, while macro scattering off of ordinary matter is purely elastic.

To estimate
the fraction of energy
that thermalizes in the anti-macro, we
consider
$p\bar{p}$ annihilation.
This 
predominantly results in multi-pion states,
with
$43\%$ of the final states including 2 charged pions and 
$49\%$ including 4 charged pions \cite{pion1},
for an average charged pion multiplicity of 3, 
and a neutral pion multiplicity of 2 \cite{pion2}.
The charged pion lifetime is $3\times 10^{-8}$s, while the neutral pion lifetime is $10^{-16}$s \cite{a}.  
Thus, all of the neutral pions decay essentially where they were produced into 2 high energy
gamma rays \cite{a},
while the charged pions or their decay products are likely to escape the macro.
% Meanwhile the charged pions almost all decay to an (anti) muon and its (anti)neutrinos -- which have long stopping lengths.
We therefore expect that 4 gamma ray photons carrying $\sim 100\,$
MeV are produced per annihilation interaction.
Following \cite{AQN8}, such photons are expected
to thermalize within the anti-macro, resulting
in an emission temperature of $T_{surf} = 10^7\,$K
from the non-degenerate part of the positron atmosphere,
with the emission spectrum expected to be strongly peaked
at $1\,$keV energies. 
Thus, $\kappa \sim 0.4$.
\iffalse
The rate of events
depends on the speed of the macro.
For definiteness, we will assume anti-macros possess a Maxwellian velocity distribution
\begin{equation}
	\label{maxwellian}
	f_{MB}(v_X) = 
		\left( \frac{1}{\pi v_{vir}^2}\right)^{\frac{3}{2}}
		4\pi v_X^2 e^{-\left(\frac{v_X}{v_{vir}}\right)^2}, 
\end{equation}
where $v_{vir} \approx 250~ \,$km$\,$s$^{-1}$.
This is the distribution of macro velocities in a non-orbiting frame moving with the Galaxy.
When considering the velocity of macros impacting a detector, e.g. on Earth,
this distribution is slightly modified by the  
motion of the Earth around the Sun
and the Sun around the galactic center.
When considering
a macro impacting a compact object
such as a white dwarf or neutron star,
the velocity of the impacting macros are
generally determined by the escape velocity
at the surface of the compact object
and any effect of the velocity of the macro
far away from the compact object
is irrelevant. 
\fi

The preceding arguments are relevant to anti-macros
that are able to travel through
the overburden of a detector
and leave a detectable
signal within it. 
The speed of an anti-macro traveling through
a medium can be determined
from Newton's second law 
\begin{equation}\label{newton}
    M_x \frac{dv_x}{dt} = \kappa \rho \sigma_x
    c^2\,.
\end{equation}
In the absence of any accelerating forces
Equation \eqref{newton} evolves as
\begin{equation}\label{speedevolution}
v^2 = v_{X,0}^2 - 2 \kappa \frac{\sigma_x}{M_x} \langle \rho x \rangle c^2\,,
\end{equation}
where $\langle \rho x\rangle = 
\int \rho dx$ is
the column density
encountered by the anti-macro
passing through the medium.
Anti-macros 
with too high a value of $\sigma_x/M_x$
would not have been expected to encounter
a detector but rather fall vertically
reaching some terminal velocity.
Indeed for $\sigma_x/M_x \geq 10^{-12}\,$
g cm$^2$ an anti-macro is not expected to reach
far below the surface of the Earth, while for
 $\sigma_x/M_x \geq 10^{-9}\,$
g cm$^2$, the anti-macro is not expected
to penetrate to the bottom of the atmosphere
with any of its initial kinetic energy.

The rest of this paper is outlined as follows.
In Section II, we constrain a wide region
of parameter space by requiring that the energy deposited by anti-macros in the early Universe not alter the CMB significantly. 
In Section III, we perform a similar calculation for big bang nucleosynthesis (BBN), where the main constraint in this case is the requirement that annihilation with helium-4 not overproduce lighter elements.
In Section IV, 
we place constraints for anti-macros that would
have caused unexpected deaths in the well-monitored population of the Western world over the past decade. In section V, we discuss constraints on 
anti-macros that are of a
similar order-of-magnitude 
to that of regular macros. 
%We discuss the relevance
%of these constraints to the AQN model
%and conclude in Section IV.

\section{Cosmic Microwave Background Constraints}
WIMPs annihilating and dumping energy into the
photon-baryon fluid would drastically alter
the CMB, leading to changes in both the
temperature and polarization power spectra. As such,
CMB anisotropies
offer an opportunity to constrain the nature of 
dark matter. Constraints
have been placed on the thermally averaged cross section
of WIMPS based
on the observed spectrum \cite{1502.01589}.
We will use this result to
constrain anti-macros annihilating with protons
in the pre-recombination fluid. 
The way in which dark matter annihilations heat the
fluid depends on the nature of the cascade of particles 
produced following the annihilations. The
fraction of the rest mass energy that is injected into the gas 
can be
modelled by an efficiency factor, $f(z)$. Computations
for various channels can be found in \cite{slatyer}. 
For all cases considered in \cite{slatyer}, 
$f(1100) \sim 0.3$ and in some cases it is closer to unity. 
As discussed earlier, in the case of anti-macros,
the fraction of energy that we expect to contribute
to the heating of the surrounding medium
is $\kappa \sim 0.4$. Thus,
we neglect a detailed calculation and take
both values to be equal to each other for simplicity.

Anti-macros would consist of a reasonable fraction
of anti-protons that would annihilate with incident
protons in the pre-recombination fluid. For WIMPS, 
which were generically considered in placing
the bounds in reference \cite{1502.01589},
the energy density
injection rate is
\begin{equation}
\dot{\rho}_{\chi} = n_\chi^2 \langle \sigma_{\chi} v \rangle
(2 m_B c^2)\,,
\end{equation}
The analogous quantity for macros is 
\begin{equation}
\dot{\rho}_{x} = n_{x} n_B \langle \sigma_{x} v \rangle
(2 m_B c^2)\,.
\end{equation}
By equating the two energy injection expressions
and utilizing the bounds from \cite{1502.01589},
we can determine the constrained
region for anti-macros. This
bound can be expressed as
\begin{equation}
    \frac{\sigma_x}{M_x} ~ < ~ 2 \times 10^{-10} \frac{cm^2}{g}\,.
\end{equation}
Macros above this bound, plotted in grey in Figure 1,
would have deposited too much energy in the early
Universe and altered the observed CMB
spectra and are thus ruled out
as being all the dark matter.

\section{Big Bang Nucleosynthesis Constraints}

The effects of antimatter injection on BBN have long been a topic of study \cite{Khlopov,Lindley,Yepes,Rehm1,K-S1,K-S2,Rehm2}.
Anti-macros can affect BBN in several different ways.  (See, e.g., Ref. \cite{Rehm2} for a detailed discussion).  They can annihilate with free protons and neutrons prior to BBN ($T > 10^{10} K$), and they can annihilate with bound nucleons inside nuclei after BBN ($T < 10^9$ K).  Photons or other particles from the annihilation process can themselves interact with (and fission) nuclei, and the light nuclei resulting from these fission and annihilation processes can yield alternative nucleosynthetic pathways.  Here we will attempt only a rough order-of-magnitude estimate of these constraints.

We will derive the limit that can be placed on anti-macros from the annihilation of the anti-macro 
with a proton bound in a $^4$He nucleus,
\begin{equation}
\label{3He}
X + ^4{\rm He} \rightarrow ^3{\rm He} + \gamma,
\end{equation}
	along with the requirement that $^3$He not be overproduced by this process.  We will not consider the additional photofission of $^4$He from annihilation-produced photons because the emission spectrum around this epoch will peak in
	the several hundred keV range, which is far below the scales of $\sim$ 10 MeV
needed to fission $^4$He. The tail of the distribution may be important and this will be the subject of a follow up study. As such, we also do not consider the production of $^6$Li from this $^3$He; both of these processes are discussed in detail in Ref. \cite{Rehm2}.
Because the $^4$He abundance produced by BBN is $\sim 10^5$ times the BBN $^3$He abundance, even a tiny fraction of destroyed $^4$He can be ruled out.

The number density of $^3$He nuclei, relative to hydrogen, produced by the process in Eq. (\ref{3He}) at time $t$ is given approximately by
\begin{equation}
(^3{\rm He/H}) = (^4{\rm He/H})~ n_x \langle \sigma_x v \rangle t.
\end{equation}
Because the macros are much more massive (and therefore moving more slowly) than the helium-4 nuclei, we
can set $v \sim \sqrt{kT/m_{^4{He}}}$.
We also have $^4{\rm (He/H)} \approx 1/12$ and
$n_x = \rho_{DM}/M_x$, where we will assume in this case that the anti-macros make up all of the dark matter
($\Omega_x \approx 0.25$).  Then we have
$n_x \approx 1.0 \times 10^{-31} {\rm cm^{-3}} ~ T^3/M_x~ {\rm g/K^3}$.  The time $t$ is related to the temperature $T$ during the epoch shortly after BBN by the relation $t = 1.8 \times 10^{20} {\rm sec}(K/T)^2$.
Combining all of these, we obtain
\begin{equation}
\label{production}
(^3{\rm He/H}) = 6.8 \times 10^{-9}~ T^{3/2} \sigma_x/M_x {\rm ~g~ cm^{-2}~ K^{-3/2}}.
\end{equation}
Eq. (\ref{production}) gives roughly the helium-3 production produced by anti-macro annihilation on helium-4 at the temperature $T$.  To get the best constraint, we set $T$ to be the temperature at which BBN terminates, $T \approx 8 \times 10^8$ K, giving
\begin{equation}
\label{production2}
(^3{\rm He/H}) = 1.5 \times 10^5 (\sigma_x/M_x){\rm ~g~ cm^{-2}}
\end{equation}
The primordial abundance of helium-3 is poorly understood; Ref. \cite{K-S1} uses the constraint
$(^3{\rm He/H}) < 3 \times 10^{-5}$, while
\cite{Rehm2} takes $(^3{\rm He/H})$ to be less than the primordial deuterium abundance (D/H) $\approx 2.6 \times 10^{-5}$ \cite{FOYY}.  Substituting the CMB bound
into Eq. (\ref{production2}) gives
$(^3{\rm He/H}) < 3.0 \times 10^{-5}$, which suggests that the CMB and BBN give similar constraints.  However, given the crudeness of the current calculation, the CMB limit is more trustworthy.  It is possible that a more detailed calculation utilizing a numerical calculation of the primordial element abundances could yield a tighter limit than our CMB constraint, but we would not expect an order of magnitude difference.  This will be checked in a future study.

\section{Human Detectors}
For a range of regular macro masses and cross sections, collisions
with the human population would have caused a detectable number
of serious injuries and deaths with obvious and unusual features, while
there have been no reports of such injuries and deaths in
regions of the world in which the human population is well-monitored.
The region of parameter
space where macros would have produced a devastating
injury similar to a gunshot wound on the carefully monitored population of the Western world was thus
constrained \cite{dbdm}. We use this same null result to constrain
the same range of anti-macro masses, which deposit
significantly more energy in human tissue.

At $1\,$keV energies, the
photons possess a scattering length in human tissue
of roughly $10^4\,$cm \cite{Xraymass}.  
Thus, $1-\exp(-0.001) \approx 10^{-3}$
of the energy from
the emission at the surface of the anti-macro is deposited in a cylinder of radius $10\,$cm.
To determine the total energy deposited, we multiply $dE/dx$ in Eq. (\ref{dedx})
by the path length of the macro inside the human body, which we assume to be $\sim 10$ cm.
The energy deposited in this radius of length 
10 cm is 
\begin{equation}
\Delta E = 10 \alpha \kappa \rho \sigma_x c^2 \,,
\end{equation}
where $\alpha = 0.001\,$ is the fraction of energy that remains
in the human tissue
through the $1\,$keV photons that interact within $10\,$cm
of the anti-macro surface and deposit
their energy in the human tissue
and $\kappa = 0.4$ is the fraction of energy
that thermalizes and is eventually deposited into the
human tissue.
Requiring $\Delta E \geq 100\,$J $=10^9\,$ergs,
a bound used previously \cite{dbdm} to constrain
regular macros, we can rule out
$\sigma_x \geq 10^{-12}\,$cm$^2$ 
% upto 
% $M_x = 5\times10^4\,$g for anti-macros.

To constrain $M_x$, we consider
the number
of encounters between an anti-macro
and the total number of humans in our sample 
\begin{equation}\label{events}
N_{events} = f\frac{\rho_{DM}}{M_x} N A t v_{x} 
\end{equation}
where $f = \Omega_{x}/\Omega_{DM}$ is 
the maximum allowed abundance
of anti-macros that can contribute
to the dark matter energy density
in the Universe,
$\rho_{DM} = 5 \times 
10^{-25}$gcm$^{-3}$\cite{n}
and we have considered a monochromatic
distribution of anti-macros with $n_x = \rho_{DM}/M_x\,$,
$N = 8\times 10^8$ is the total number of humans
in our sample
$A \sim 1\,$m$^2$ is the
cross-sectional area of a human, 
$t = 10\,$years is the exposure time of our detector 
and $v_x$ is the (relative) 
speed between the anti-macro and a human. (For more details, see the corresponding discussion in Ref. \cite{dbdm}).

Since the impact of an anti-macro
is a Poisson process,
the probability $P(n)$ of 
$n$ impacts over the exposure time $t$
follows the Poisson distribution
\begin{equation}
\label{eq:Poisson}
P(n) = \frac{{N_{events}}^n}{n!}
e^{-N_{events}}\,.
\end{equation}
where $N_{events}$ is the expected
number of events per interval
and was given in Equation \eqref{events}. 
If no events are observed, then
the value $N_{events} \geq 3$ may be ruled
out at 95\% confidence, i.e.
by requiring that the 
probability of no detected signals be
less than $5 \%$.
This allows us to constrain
the abundance of anti-macros
as a function of the mass $M_x$
\begin{equation}
f \leq \frac{M_x}{5\times 10^4 g}\,.
\end{equation}

\section{Other constraints}

For many physical processes, the limits that can be placed on anti-macros are identical to the corresponding limits on macros.  We discuss those limits briefly here.

\subsection{Microlensing}
{\bob For very large masses} ($M_x \geq 10^{23}\,$g), a variety of microlensing searches have  constrained 
heavy composite object candidates
\cite{i2,j,k,l,m} to make up at most
a sub-leading component of the dark matter, regardless
the nature of their non-gravitational interactions.

\subsection{Paleo-detectors}
Macros that would have penetrated 
a few km into the Earth’s crust would have 
left tracks in ancient muscovite mica. Searches for grand-unified theory magnetic monopoles \cite{h,i} sought to detect lattice defects left in ancient mica
through chemical etching techniques.
These limits were used to place
limits on regular macros,
over a wide range of cross-sections,
up to $M_x = 55\,$g \cite{b} (this idea of using paleo-detectors
to detect dark matter is not new; see \cite{Edwards,Drukier}
for an approach to detect WIMPs with paleo-detectors). 
% In order to have left an etchable track, the anti-macro
% must have deposited 
% a minimum nuclear component of stopping power 
% \begin{equation}
% \frac{dE}{dx} \sim 10 \frac{GeV}{cm}. 
% \end{equation}
We expect the elastic scattering cross-section
of such objects to be of the same
order as the annihilation cross-section for
anti-macros \cite{Klempt}\footnote{No data
exists to the best of our knowledge for 
antiproton energies of less than $\sim 100\,$
MeV. However, we expect that
the order 1 difference between the annihilation
and elastic scattering cross-sections
at $100\,$MeV to not increase by several
orders of magnitude at energies
of several hundred keV.} and so the
same results apply as in the case of regular macros.

We have also 
suggested that, for appropriate $M_x$ and $\sigma_x$, the passage of a macro through granite
would form long tracks of melted and re-solidified rock
that would be distinguishable from the surrounding unmelted granite \cite{q}. A search for such tracks in commercially
available granite slabs is planned.
Using the same reasoning that allowed
us to use the null result from the etching
of ancient muscovite mica, anti-macros of 
the same minimum cross-section would
have left visible tracks in the granite
and so we expect the same results of hold.
Thus, the results from the search of
slabs of mica will apply to anti-macros as well.

\subsection{Thermonuclear runaway}
As
discussed in \cite{thermo1} (and references therein), for thermonuclear runaway to be ignited, there is a minimum sized region, $\lambda_{trig}$ that must
be raised above a threshold temperature $T_{crit} \sim
3 \times 10^9\,$K, where $\lambda_{trig}$ is
strongly dependent on density.
Constraints were placed on elastically
scattering macros using the continued existence of
white dwarfs \cite{thermo1}. However, it was later
determined that these constraints were too stringent;
more accurate bounds were placed \cite{thermo2} although
these bounds are subject to additional uncertainties.
It has not been confirmed through numerical
simulations
that the conditions identified in \cite{thermo2,thermo1} 
are indeed sufficient
to initiate thermonuclear runaway, i.e., there remains
some uncertainty whether in fact heating a region of size
at least $\lambda_{trig}$ to $T \sim$ few $\times 10^9\,$K 
necessarily causes type
1A supernovae in white dwarfs and superbursts in neutron stars. 
In the case of anti-macros, the energy deposition
is not
expected to be
much higher than the case of macros. 
This is because the emission
temperature of the anti-macro is expected
to be high enough that most of the positrons
in the non-degenerate regime will be ionized.\footnote {The temperature is higher
in this case due to the larger number density of the target medium; see appendix A in reference \cite{AQN8}.} Thus, $\lambda_{trig}$ is
still determined primarily by the cross-section of the
anti-macro
similar to the case of regular macros \cite{thermo2}. 
Emission from denser regions near the core
is highly suppressed (see Appendix 4 in \cite{AQN2}).

\section{Discussion and Conclusions}
We have considered a phenomenological approach
and constrained the abundance of anti-macros over the relevant mass range
based on several terrestrial, astrophysical and
cosmological probes. Atomic density anti-macros are
entirely ruled out by a combination of the CMB
and microlensing constraints. Nuclear density macros
are ruled out below $5\times 10^4\,$g, and possibly
at some higher range mass windows through thermonuclear 
runaway. However those results are subject to additional
uncertainties as discussed in Section IV.
% The results presented here constrain macros

% We note that one may expect such antimatter objects
% to be visible in the form of meteors or bolides
% passing through the atmosphere. Indeed,
% constraints were placed using
% networks of cameras that were originally built to study
% bolides, i.e. extremely bright meteorites \cite{1908.00557}.
% The constraints were placed by using the null
% observation of macro-induced bolides capable
% of illuminating the pixels of the camera at a flux
% greater than the threshold flux.
% However, as shown in \cite{AQN}, the emission 
% spectrum is strongly peaked above the
% visible light portion of the electromagnetic spectrum
% although the energy deposited is much greater in
% the case of anti-macros compared to regular macros, 
% by a factor
% of $\sim 10^6\,.$ Furthermore, the mean free paths
% of electrons ionized by the high energy photons
% is expected to be larger than the
% intermolecular distance resulting in the formation of a 
% diffuse ionized gas. This diffuse gas will have a much
% longer recombination time $t_{rec} \propto n_e^{-1}\,,$
% than in the case where a strongly ionized plasma was
% formed with elastically scattering macros. Additionally, 
% the trajectory of the anti-macro will be difficult to
% ascertain since
% the recombination across the
% entire field of view will likely be random due
% to the large distances travelled by the photons
% and electrons produced. This will likely result in an
% unfruitful search
% using bolide detecting cameras, and the much more
% sensitive Fluorescence Detectors (FDs). 

We also note that such anti-matter objects could in
principle alter the reionization history, as discussed
in \cite{1409.5736}. However, the most
stringent constraints come from the early Universe due to
the large number density of both the anti-macros
and the protons. For objects below the grey bound in
Figure 1, which are significantly denser than atomic
density objects, 
the number of encounters of an anti-macro
with hydrogen atoms was found to be low enough that it
will not significantly alter the
recombination history nor produce
a higher extragalactic photon background around the 100 MeV
range.

% The results from the lack of gruesome
% and unexplained deaths are particularly constraining
% for the
% axion quark nugget (AQN) dark matter model. Those
% results constrain $\langle B \rangle \leq 3\times 10^{28}\,$
% ruling out the entire preferred region for AQNs \cite{AQN8}.
% However, we note that the seismological constraints
% inferred in \cite{AQN8} should not be trusted
% as shown in \cite{seismo}, where a more
% careful ray tracing algorithm was performed
% and the results in \cite{teplitz} considerably
% weakened. Thus, the search for AQNs, and anti-macros
% in general must continue.

% We have ignored the fraction of energy lost to, e.g.
% axions, which could be a significant fraction
% of the energy that is emitted at the surface of these
% AQNs. However, this is unlikely to significantly
% alter the results of this section.
% The temperature at the surface of the anti-macro/AQN
% varies weakly (see Appendix A of reference \cite{AQN}
% with this fraction of energy that is lost
% through weakly interacting particles such as axions
% and neutrinos. Secondly, 1/3 of the mass of the 
% AQN is the energy associated with the axion field and 
% not related to the baryon charge. Thus, this is not
% expected to alter our constraint significantly.

\begin{figure*}[h]
\centering
\includegraphics[width=\textwidth]{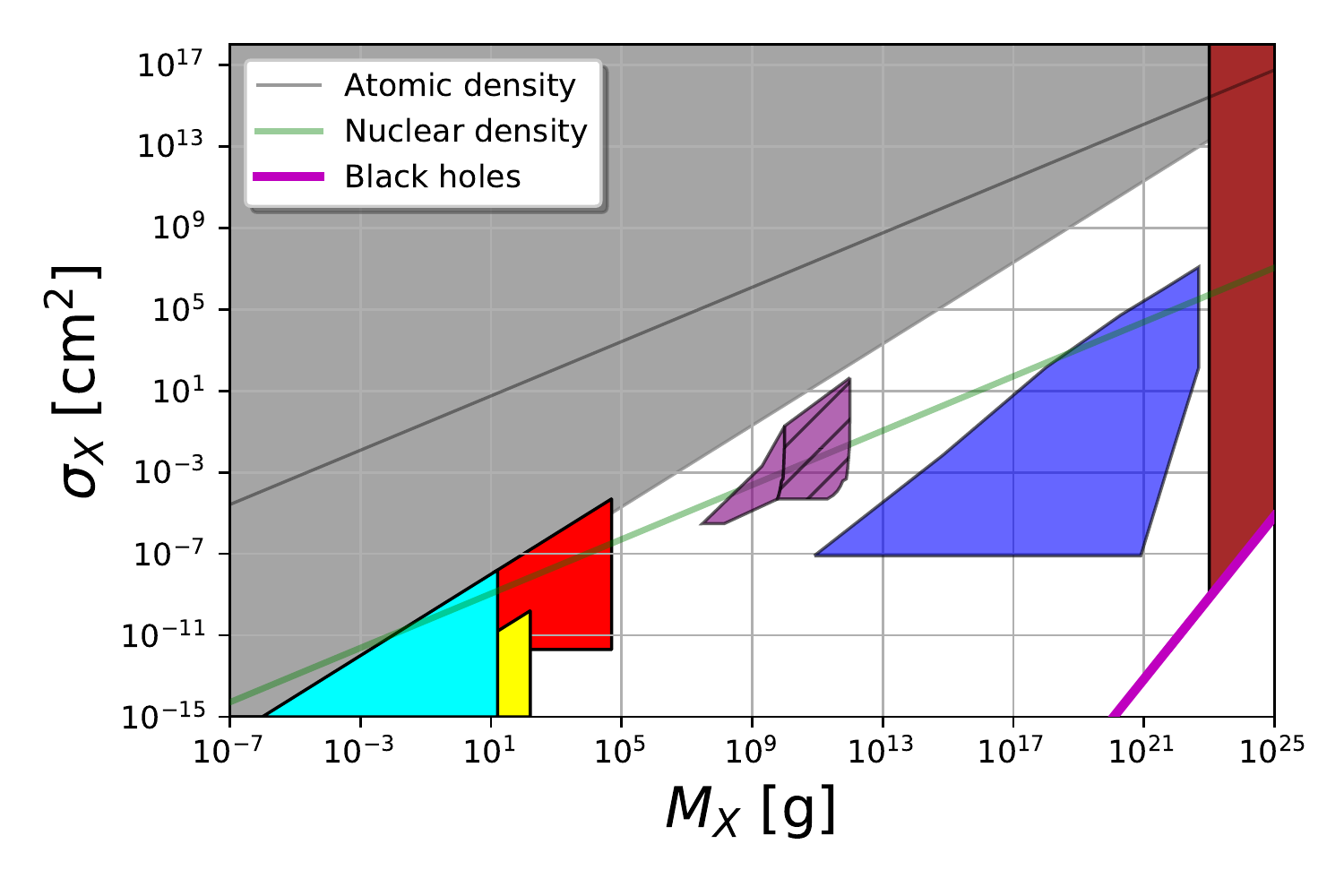}
\caption{Constraints on the anti-macro cross section and mass (assuming the anti-macros have a single mass).
Constraints in brown
come from various microlensing experiments, in 
yellow from a lack of tracks in an
ancient slab of mica \cite{h,i}, in cyan from
a lack of a signal in the Icecube experiment \cite{AQN8};
in grey 
from the Planck Cosmic Microwave Background,
in red from a lack of human impacts, in light blue
from thermonuclear runaway in white dwarfs
and in light purple from a lack of superbursts
over a period of a decade on the
superburster 4U 1820-30. The hatched light purple
region shows the region of parameter space that may be
constrained from an analysis of all known local 
X-ray binaries and a better understanding of the mean
background superburst rate \cite{thermo2}.
The black and green lines
correspond to objects of constant density
1 g cm$^{-3}$ 
and
$3.6 \times 10^{14}$ g cm$^{-3}$
respectively. Black hole candidates lie on the magenta line.}
\label{fig:dbdm}
\end{figure*}

\section*{Acknowledgements} 
This work was partially supported by Department of
Energy grant de-sc0009946 to the particle astrophysics
theory group at CWRU. 
R.J.S. was partially supported by the  Department of Energy,
de-sc0019207.


\begin{thebibliography}{99}

\bibitem{a}

	Tanabashi, M. and others,
    \emph{Review of Particle Physics},
    \emph{Phys. Rev. D}, {\bf 98},
    {030001}, {2018}

\bibitem{b}
	David M. Jacobs and Glenn D. Starkman and Bryan W. Lynn,
	\emph{Macro dark matter},
	\emph{Monthly Notices of the Royal Astronomical Society},
	{\bf 450}, {3418 -- 3430}, {2015}
	
\bibitem{b1}
	David M. Jacobs and Amanda Weltman and Glenn D. Starkman,
	\emph{Resonant bar detector constraints on macro dark matter},
	\emph{Phys. Rev. D},
	{\bf 91}, {115023}, {2015}
	
\bibitem{atmos}
Sidhu, Jagjit Singh and Abraham, Roshan Mammen and
                        Covault, Corbin and Starkman, Glenn,
                        \emph{Macro detection using fluorescence detectors},
                        \emph{JCAP}, {\bf 1902}, {037}, {2019}.
                        
\bibitem{q}
Jagjit Singh Sidhu and Glenn Starkman and Ralph Harvey,
\emph{A counter-top search for macroscopic dark matter},
\emph{arXiv:1905.10025}, {2019}

\bibitem{dbdm}
Jagjit Singh Sidhu, Robert Scherrer, Glenn Starkman,
\emph{Death and Serious Injury by Dark Matter},
\emph{Physics Letters B}, {\bf 803}, {135300}, {2020}

\bibitem{1908.00557}
Jagjit Singh Sidhu and Glenn Starkman, \emph{Macroscopic Dark Matter Constraints from Bolide Camera Networks}, \emph{Phys. Rev. D}
{\bf 100}, {123008}, {2019}

\bibitem{thermo2}
Jagjit Singh Sidhu and Glenn Starkman, \emph{Reconsidering
Astrophysical Constraints on macroscopic dark matter}, \emph{Phys. Rev. D}
{\bf 101}, {083503}, {2020}

\bibitem{g2}
    Jagjit Singh Sidhu,
    \emph{Charge constraints of macroscopic dark matter},
    \emph{Phys. Rev. D},
    {\bf 101}{043526}, {2020}
    
\bibitem{h}

	De Rujula, A. and Glashow, S. L.,
    \emph{Nuclearites: A Novel Form of Cosmic Radiation},
    \emph{Nature}, {\bf 312}, {734-737}, {1984}

\bibitem{i}
	Price, P. B.,
    \emph{Limits on Contribution of Cosmic Nuclearites to Galactic
                        Dark Matter},
    \emph{Phys. Rev. D},
    {\bf 38},
    {3813-3814}, {1988}.
    
\bibitem{i2}
	H. Niikura et. al,
    \emph{Microlensing constraints on primordial black holes with Subaru/HSC Andromeda observations},
    \emph{Nature Astronomy},
    {\bf 3},
    {524-534}, {2019}.

\bibitem{j}
P. Tisserand and others, \emph{Limits on the Macho content of the Galactic Halo from the {EROS}-2 Survey of the Magellanic Clouds}, \emph{Astronomy {\&} Astrophysics},
{\bf 469}, {387--404}, {2007}

\bibitem{k}
C. Alcock and others, \emph{{MACHO} Project Limits on Black Hole Dark Matter in the 1{\textendash}30 $M_\odot$ Range},
\emph{The Astrophysical Journal},
{\bf 550}, {L169-L172}, {2001}.

\bibitem{l}
B. J. Carr and Kazunori Kohri and Yuuiti Sendouda and Jun'ichi Yokoyama,
\emph{New cosmological constraints on primordial black holes},
\emph{Phys. Rev. D}, {\bf 81}, {104019}, {2010}

\bibitem{m}
Kim Griest and Agnieszka M. Cieplak and Matthew J. Lehner, 
\emph{New Limits on Primordial Black Hole Dark Matter from an Analysis of Kepler Source Microlensing Data}, 
\emph{Physical Review Letters}, {\bf 111}, {181302}, {2013}

\bibitem{thermo1}
Peter W. Graham and Ryan Janish and Vijay Narayan and Surjeet Rajendran and Paul Riggins, \emph{White dwarfs as dark matter detectors}, \emph{Phys. Rev. D}
{\bf 11}, {115027}, {2018}

\bibitem{Boehm}
Ryan J. Wilkinson and Julien Lesgourgues and Celine Boehm,
\emph{Using the CMB angular power spectrum to study Dark Matter-photon interactions}, \emph{JCAP}, {\bf 2013},
{026}, (2013)


\bibitem{c}

	Witten E, \emph{Cosmic Separation of Phases,}, \emph{Phys. Rev. D} {\bf 30}{272-285} (1984)

% \bibitem{d}

% 	Bryan W. Lynn and Ann E. Nelson and Nikolaos Tetradis, \emph{Strange Baryon Matter,} \emph{Nuc. Phys. B} {\bf 345:}{186-209} (1990)

% \bibitem{e}

% 	Bryan W. Lynn, \emph{Liquid Phases in {SU(3)} {C}hiral {P}erturbation {T}heory: {D}rops of 
% {S}trange {C}hiral {N}ucleon {L}iquid and {O}rdinary {C}hiral {H}eavy {N}uclear 
% {L}iquid}, \emph{arXiv:1005.2124}, {2010}


% \bibitem{f}

% 	Nelson, Ann E.,
%     \emph{Kaon Condensation in the Early Universe},
%      \emph{Physical Letters},
%     {\bf B240}, {179-182}, {1990}

\bibitem{AQN1}

	Ariel R. Zhitnitsky, \emph{"Nonbaryonic" Dark Matter as Baryonic Color Superconductor}, \emph{JCAP} 
	{\bf 2003} {010} (2003)
	
\bibitem{AQN2}

    Michael M Forbes and Ariel R. Zhitnitsky, \emph{WMAP haze: Directly observing dark matter?}, \emph{Phys. Rev. D} 
	{\bf 78} {083505}, (2008)
	
\bibitem{AQN3}

    	Xunyu Liang and Ariel R. Zhitnitsky, \emph{Axion field and the quark nugget's formation at the QCD phase transition}, \emph{Phys. Rev. D} 
	{\bf 94} {083502} (2016)


	
\bibitem{AQN4}
    	Shuailiang Ge and Xunyu Liang and Ariel Zhitnitsky, \emph{Cosmological Axion and Quark Nugget Dark Matter Model}, \emph{Phys. Rev. D} 
	{\bf 97} {043008} (2018)

	
\bibitem{AQN5}

    	Nayyer Raza and Ludovic van Waerbeke and Ariel Zhitnitsky, \emph{Solar Corona Heating by the Axion Quark Nugget Dark Matter}, \emph{Phys. Rev. D} 
	{\bf 98} {103527}, (2018)

	
\bibitem{AQN6}
    
    
	Victor V. Flambaum and Ariel R. Zhitnitsky, \emph{Primordial Lithium Puzzle and the Axion Quark Nugget Dark Matter Model}, \emph{Phys. Rev. D} 
	{\bf 99} {023517} (2019)

	
\bibitem{AQN7}

    	Shuailiang Ge and Kyle Lawson and Ariel Zhitnitsky, \emph{The Axion Quark Nugget Dark Matter Model: Size Distribution and Survival Pattern}, \emph{Phys. Rev. D} 
	{\bf 99} {116017} (2019)

	
\bibitem{AQN8}

Dmitry Budker and Victor V. Flambaum and Ariel Zhitnitsky,
\emph{Axion Quark Nuggets. SkyQuakes and Other Mysterious Explosions}, \emph{arXiv: 2003.07363} {2020}

\bibitem{1502.01589}
Planck Collaboration and P. A. R. Ade and 
others,
\emph{Planck 2015 results. XIII. Cosmological parameters},
Year = {2015},
\emph{{A$\&$A}}, {\bf 594}, {A13}, (2016)

\bibitem{slatyer}
  Tracy R. Slatyer and Nikhil Padmanabhan and Douglas P. Finkbeiner, \emph{{CMB} constraints on {WIMP} annihilation: Energy absorption during the recombination epoch},
  \emph{Phys. Rev. D.}, {\bf 80}, {043526}, (2015)

\bibitem{pion1}
Claude Amsler, \emph{Nucleon-antinucleon annihilation at LEAR}, \emph{arXiv:1908.08455}, {2019}.

\bibitem{pion2}
Claude Amsler, \emph{Proton-antiproton annihilation and meson spectroscopy with the Crystal Barrel}, \emph{Review of
Modern Physics}, {\bf 70}, {1293}, {1998}.

\bibitem{Khlopov}
V.M. Chechetkin, M.Yu. Khlopov, M.G. Sapozhnikov,
and Ya. B. Zeldovich, \emph{Astrophysical aspects of antiproton interaction with $^4$He (antimatter in the universe)}, \emph{Phys. Lett. B}, {\bf 118}, 329 (1982).

\bibitem{Lindley}
David Lindley, \emph{Hadronic decays of cosmological gravitinos}, \emph{Phys. Lett. B}, {\bf 171}, 235 (1986)

\bibitem{Yepes}
G. Yepes and R. Dominguez-Tenreiro,
\emph{The effects of antimatter on primordial nucleosynthesis},
\emph{Ap. J.}, {\bf 335}, 3 (1988).

\bibitem{Rehm1}
Jan B. Rehm and Karsten Jedamzik, \emph{Big bang nucleosynthesis with matter-antimatter domains},
\emph{Phys. Rev. Lett.}, {\bf 81}, 3307 (1998).

\bibitem{K-S1}
H. Kurki-Suonia and E. Sihvola,
\emph{Constraining antimatter domains in the early universe with big bang nucleosynthesis}

\bibitem{K-S2}
H. Kurki-Suonio and E. Sihvola,
\emph{Antimatter regions in the early universe and big bang nucleosynthesis}, \emph{Phys. Rev. D}, {\bf 62}, 103508 (2000).

\bibitem{Rehm2}
Jan B. Rehm and Karsten Jedamzik, \emph{Limits on Matter-Antimatter Domains from Big Bang Nucleosynthesis,
} \emph{Physical Review D}, {\bf 63}, 043509 (2001).

\bibitem{FOYY}
Brian D. Fields, Keith A. Olive, Tsung-Han Yeh, and Charles Young, \emph{Big-bang nucleosynthesis after Planck}, \emph{JCAP}, {\bf 3}, 010 (2020).

\bibitem{Xraymass}
Stephen Seltzer, \emph{Tables of X-Ray Mass Attenuation Coefficients and Mass Energy-Absorption
              Coefficients,  NIST Standard Reference Database 126}, \emph{National Institute of Standards and Technology}
              
              
\bibitem{n}
Jo Bovy and Scott Tremaine, \emph{{O}n THE LOCAL DARK MATTER DENSITY},
\emph{The Astrophysical Journal}, {\bf 756}, {89}, {2012}
             
\bibitem{Edwards}
	Edwards, Thomas D. P. and Kavanagh, Bradley J. and
                        Weniger, Christoph and Baum, Sebastian and Drukier,
                        Andrzej K. and Freese, Katherine and Górski, Maciej and
                        Stengel, Patrick,
    \emph{Digging for dark matter: Spectral analysis and discovery
                        potential of paleo-detectors},
    \emph{Phys. Rev. D}, {\bf 99}, {043541}, {2019}
    
\bibitem{Drukier}
	Drukier, Andrzej K. and Baum, Sebastian and Freese,
                        Katherine and Górski, Maciej and Stengel, Patrick
    \emph{Paleo-detectors: Searching for Dark Matter with Ancient
                        Minerals},
    \emph{Phys. Rev. D}, {\bf 99}, {043014}, {2019}
    


\bibitem{Klempt}
Eberhard Klempt and Franco Bradamante and Anna Martin and Jean-Marc Richard, \emph{Antinucleon{\textendash}nucleon interaction at low energy: scattering and protonium},
\emph{Elsevier {BV}}, {\bf 368}, {119-316}, {2002}.

\bibitem{1409.5736}
S.I. Blinnikov and A.D. Dolgov and K. A. Postnov, \emph{Antimatter and antistars in the universe and in the Galaxy}, \emph{Phys. Rev. D}
{\bf 92}, {023516}, {2015}


\end{thebibliography}
\end{document}